\def\np{Nucl. Phys.}
\def\pl{Phys. Lett.}
\def\prl{Phys. Rev. Lett.}
\def\pr{Phys. Rev.}

 \def\be{\begin{equation}}
\def\ee{\end{equation}}
\def\ba{\begin{eqnarray}}
\def\ea{\end{eqnarray}}
\def\sw{\mbox{$\sin^2\theta_w$}}
\def\Sw{\sin^2\theta_w}
\def\slash{\hspace{-5pt}/}
\def\el{\mbox{$e_L$}}
\def\elbar{\mbox{$\bar e_L$}}
\def\blrbar{\mbox{$\bar b_{L,R}$}}
\def\blr{\mbox{$b_{L,R}$}}
\documentstyle[12pt]{article}
\begin{document}
\title{Reconciling the LEP and SLAC measurements of \sw}
\author{Francesco Caravaglios$^a$\thanks{INFN Fellow},   \\Graham G.
Ross$^b$\thanks{SERC
Senior
Fellow, on leave from $^a$},\\
\\$^a $Department of Physics,
Theoretical Physics,\\
University of Oxford,
1 Keble Road,
Oxford OX1 3NP\\ \\ $^b$ Theory Division, CERN, CH-1211, Geneva, Switzerland}
\date{}
\maketitle

\begin{abstract}
We consider whether a  discrepancy between the SLAC and LEP measurements of
$\Sw$ can be explained by new physics. We find that  only the contribution of a
new
neutral gauge boson, $Z^\prime$, nearly degenerate with the Z  can affect the
SLAC
measurement while leaving the LEP observables almost unaffected.  We briefly
discuss
possible signals for this new gauge boson, including changes in the $Z$
lineshape when measured with polarised electrons, small changes in $R_b$,
$A^e_{FB}$, and larger changes in two jet and $t\bar{t}$ production  at hadron
colliders.
\end{abstract}



\vspace{-16cm}\hspace{12cm} CERN-TH.7474/94 \\

\hspace{12cm} OUTP 9424P
\vspace{15cm}
\section*{Introduction}

In the context of the Standard Model, the value of \sw determined by SLAC
\cite{slac}
from the measurement of the $A_{LR}$ asymmetry currently disagrees at the 2.5
standard deviation level with the value obtained from a variety of precision
measurements performed at the LEP collider.  All channels at the LEP
experiments\cite{glasgow} give a value of the Weinberg angle
($\sin^2\theta_w=0.2321\pm 0.0004$)  which is consistent with the Standard
Model
prediction \cite{bardin}\cite{altlampe}\cite{veltman} ($\sin^2\theta_w=0.2320$)
for a top
mass of about $174~GeV$ \cite{cdf}. On the other hand the $A_{LR}$ asymmetry
measured at
SLAC gives $\Sw=0.2292\pm 0.001$ requiring a much heavier top quark for
consistency
with the Standard Model.

The immediate question raised by this discrepancy is whether it signals new
physics
beyond the Standard Model. In this letter we will discuss the nature of the new
physics
that can allow the two measurements to be consistent. We will show that the
only
possibility is the existence of a new neutral gauge boson, $Z^\prime$, whose
mass and
coupling to the fermions is strongly constrained. Moreover this new gauge boson
may
be responsible for the small excess in the LEP measurement of
$R_b=\Gamma_b/\Gamma_h=0.2192\pm 0.0018$ \cite{glasgow} compared to  the
Standard Model Prediction \cite{Akundov}\cite{bardin}
$R_b=\Gamma_b/\Gamma_h=0.2157\pm0.0005$ ($M_{top}=175\pm 15~GeV$,
$M_{H}=300~GeV$). Finally we consider possible tests for such a new gauge
boson.

\section*{Constraints on the new physics}

The possibility of explaining the difference between the LEP and SLAC
measurements
through new physics arises because they refer to different observables:
\begin{itemize}
\item SLAC uses polarized initial beams to measure the LR asymmetry
\be
\label{eq:alr}
A_{LR}={\sigma_L-\sigma_R \over \sigma_L +\sigma_R}
\ee
where $\sigma_L$ and $\sigma_R$ are respectively the cross sections (at the
$Z_0$
peak) of
\ba
\label{eq:process}
e_L+\bar e_L &\longrightarrow&  X\\
e_R+\bar e_R &\longrightarrow&  X\nonumber
\ea
where $e_{L(R)}$ is a lefthanded (righthanded) electron and $X$ is an hadronic
or a
$\tau^+\tau^-$  final state.  The value measured at SLAC is

\be
\label{eq:slac}
A_{LR}^{SLAC}=0.163\pm 0.0079
\ee

\item LEP uses unpolarized initial beams to study the
asymmetries of a fermion-antifermion pair in the
final state. Within the Standard Model we may use the value of \sw obtained at
LEP
(averaged over all channels) to  predict the LR asymmetry (henceforth we will
call
$A_{LR}^{LEP}$ the {\it prediction} of LEP measurements for the SLAC asymmetry)
\be
\label{eq:lep}
A_{LR}^{LEP}=0.142\pm 0.0032
\ee

\end{itemize}
In the Standard Model $A_{LR}^{SLAC}$ and $A_{LR}^{LEP}$ should be equal but
new physics can change the expectation for one or both. However (with the
possible
exception of $R_b$ which is discussed below), the consistency of  all LEP
precision
measurements with the Standard Model for a top mass of about $174~GeV$
\cite{cdf}
means that changing $A_{LR}^{LEP}$ significantly (via a change of \sw) is
unacceptable\footnote{A particularly clear case is the LEP measurement of
$A_{FB}^e$
: as will be obvious from eq(\ref{eq:interf}) with q=e, it is not possible to
{\it decrease }
$A_{FB}^e$ to the value $0.0156$ \cite{glasgow} obtained if one uses the SLAC
result
$g_V^e/g_A^e=0.082$.}\cite{erler}.

Thus we must look for new physics that changes $A_{LR}^{SLAC}$ while leaving
$A_{LR}^{LEP}$  LEP measurements
essentially unchanged. In particular the accuracy of the LEP measurement for
the total
unpolarised hadronic cross section (plus  $\tau^+~\tau^-$ events)
\be
\label{eq:lephad}
{\sigma_L +\sigma_R\over 2}  =(43.49\pm 0.12)~nb~~ ~~~(LEP)  \ee
is so precise (and in agreement with the Standard Model
prediction) that we must require the changes $\delta\sigma_{R,L}$ in
$\sigma_{L,R}$
to satisfy

\begin{equation}
\delta \sigma_R+\delta \sigma_L \simeq 0
\label{eq:con}
\end{equation}
to the accuracy of eq(\ref{eq:lephad}). This accuracy is much better than the
discrepancy between eqs.(\ref{eq:slac}) and (\ref{eq:lep}). Hence the
requirement that
the theoretical prediction for  $A_{LR}^{SLAC}$ be increased means that the new
physics must give $\delta\sigma_R\approx-\delta\sigma_L$  and $\delta
\sigma_{R}/\sigma_{R}\simeq -(1\div2)\%$.

The new contribution needed  to generate $\delta \sigma_R <0 $ must come from
an
interference between the Standard Model amplitude and the amplitude coming from
the
new physics because a non-interfering term would necessarly give $\delta
\sigma_R >0$.
 The Standard Model amplitude for the process of eq(\ref{eq:process}) has the
form
\be
\label{eq:m0}
M_0^{R,L}=  a \bar u(p_q) \gamma^\mu (g_V^q+g_A^q \gamma_5) v(q_q)  \times
\bar v(p_e) \gamma_\mu (g_V^e+g_A^e \gamma_5)\left({1\pm \gamma_5 \over
2}\right)
u(q_e)  \ee
corresponding to the processes (\ref{eq:process})  with
$q^2=M_Z^2$. The quantity {\it a} is determined by the Z propagator and on
resonance
is purely imaginary. In order to interfere with this amplitude the new physics
must
generate an amplitude, $\delta M^{L,R}$, of the form
\be
\label{eq:dm}
\delta M^{L,R}=\delta a\bar u(p_q) \gamma^\mu
(g_V^{q\prime}+g_A^{q\prime}\gamma_5) v(q_q)  \times
\bar v(p_e) \gamma_\mu (g_V^{e\prime}+g_A^{e\prime}
\gamma_5)\left({1\pm \gamma_5 \over 2}\right) u(q_e)
\ee
with  $\delta a$ imaginary\footnote{Neglecting real terms which do not intefere
is a
simplification which do not alter the conclusions of the following discussion
because they
are strongly constrained by the requirement that LEP measurements be
unaffected to the accuracy of eq(\ref{eq:lephad}).}. The  squared matrix
element is
\be
|M_0^{L,R}+\delta M^{R,L}|^2=|M_0^{R,L}|^2+ 2 Re(M_0^{L,R} \delta M^{{L,R}*})+
|\delta M^{L,R}|^2.
\label{eq:int}
\ee
The first term is the Standard Model contribution, and   we assume  for the
moment the
third one to be small compared with the others.  The second term is the
interference
between the $Z_0$ contribution and the new physics contribution. In the
processes
measured at LEP we must  take the average  over the initial polarizations
giving
(ignoring fermion masses)
\ba
\label{eq:interf}
 { \mbox{\Large $\Sigma$}_{L,R}} ~\mbox{\large $Re$}(M_0^{L,R} \delta
M^{{L,R}*}) &=&  \mbox{\large $Re$}( a \delta a^*Tr\left(\gamma^\mu p\slash_e
\gamma^\nu q\slash_e\left(
(g_V^e g_V^{e\prime} +g_A^e g_A^{e\prime})+ (g_V^e g_A^{e\prime} +g_V^{e\prime}
g_A^e)\gamma_5\right)\right))\nonumber \\ &&
\times
Tr\left(\gamma_\mu p\slash_q \gamma_\nu
q\slash_q\left(
(g_V^q g_V^{q\prime} +g_A^q g_A^{q\prime})+ (g_V^q g_A^{q\prime}
+g_V^{q\prime} g_A^q)\gamma_5\right)\right)
\ea
If the predictions for the LEP measurements in this channel labelled by q are
to remain
essentially those of the Standard Model we must require
\be
\label{eq:relat1}
(g_V^e g_V^{e\prime} +g_A^e g_A^{e\prime})\simeq 0 ~~~and ~~~~ (g_V^q
g_A^{q\prime} +g_A^q g_V^{q\prime})\simeq 0
\ee
or
\be
\label{eq:relat2}
 (g_V^e g_A^{e\prime} +g_A^e g_V^{e\prime})\simeq 0~~~and ~~~~   (g_V^q
g_V^{q\prime} +g_A^q g_A^{q\prime})\simeq 0.
\ee
If either of these conditions holds one of the traces in
(\ref{eq:interf}) has the term proportional  to $\gamma_5$ vanishing, while the
other has
only the term proportional  to $\gamma_5$ non-vanishing.
Thus the vanishing of the interference (\ref{eq:interf}) may be immediately
seen because
one term is symmetric in  ($\mu,\nu$) while the other is antisymmetric. Further
by definition either of the relations in eqs(\ref{eq:relat1}) or
eq(\ref{eq:relat2}) imply $\delta \sigma_L (\theta)\simeq -\delta
\sigma_R(\theta)$ where
$\theta$ is the centre of mass scattering angle. Finally we note that the
relation
(\ref{eq:relat1})  implies $\delta \sigma_R(\theta))$ is  even in
$\cos(\theta)$ while  the
relation (\ref{eq:relat2}) implies that it is odd so that only the former
allows for a non-
zero amplitude integrated over $\theta$. Putting all this together we conclude
that the
new physics must generate an amplitude  of the form of
eq(\ref{eq:dm}) with {\it a} imaginary and further constrained by
eq(\ref{eq:relat1}).
Only in this case can one change the prediction for the SLAC asymmetry
measurement
while leaving the predictions for the LEP measurements unchanged.

However this is possible only for a restricted class of final states X. In the
case X is $e^+e^-$ it is impossible to satisfy eq(\ref{eq:relat1}) and moreover
the additional non-Z contribution is positive. Thus to preserve the agreement
of the LEP measurements with the Standard Model we must keep the couplings of
the $Z'$ to the electron small. In the case $X=\tau^+\tau^-$ the consistency of
the LEP $\tau$ polarisation measurements together with the unpolarised $\tau$
measurements require eq(\ref{eq:relat1}) be satisfied for
$g_V^{q,q'}=\pm g_A^{q,q'}$, $q=\tau$ (i.e. purely left-handed and right-handed
final states). These can only be satisfied if the coupling of the new physics
to the $\tau$ is small. We note that the $\tau$ constraints are satisfied if we
assume universal lepton couplings and if the electron constraints discussed
above are satisfied.

\section*{Nature of the new physics}
We turn now to a discussion of the nature of the physics beyond the Standard
Model capable of generating such a matrix element.The discussion above implies
that the new physics is predominantly coupled to the quark sector.
We will consider the following three possibilities which arise at tree or one
loop level:

i). The new physics generates a correction to the Z coupling as in Fig 1a.

ii). The new physics generates a new box contribution as in Fig 1b.

iii). The new physics generates a new contribution at tree level via a Born
graph.
\vskip 36pt
\begin{center}
\setlength{\unitlength}{0.0075in}%
\begin{picture}(595,165)(20,645)
\thicklines
\put( 85,775){\line( 0,-1){ 70}}
\put(125,740){\line( 1, 0){100}}
\put(305,670){\line(-6, 5){ 81.639}}
\put(225,740){\line( 6, 5){ 81.639}}
\put( 65,800){\line( 0,-1){ 10}}
\put( 65,790){\line(-1, 0){ 10}}
\put(265,715){\line( 0,-1){ 10}}
\put(265,705){\line(-1, 0){ 10}}
\put(275,775){\line(-1, 0){ 10}}
\put(265,775){\line( 0, 1){ 10}}
\multiput(175,810)(0.00000,-8.00000){18}{\line( 0,-1){  4.000}}
\put( 75,690){\line(-1, 0){ 10}}
\put( 65,690){\line( 0, 1){ 10}}
\put( 45,810){\line( 6,-5){ 81.639}}
\put(125,740){\line(-6,-5){ 81.639}}
\put(505,709){\framebox(60,60){}}
\put(565,769){\line( 1, 1){ 40}}
\put(505,769){\line(-1, 1){ 40}}
\put(465,669){\line( 1, 1){ 40}}
\put(485,799){\line( 0,-1){ 10}}
\put(485,789){\line(-1, 0){ 10}}
\put(585,699){\line( 0,-1){ 10}}
\put(585,689){\line(-1, 0){ 10}}
\put(495,689){\line(-1, 0){ 10}}
\put(485,689){\line( 0, 1){ 10}}
\put(595,789){\line(-1, 0){ 10}}
\put(585,789){\line( 0, 1){ 10}}
\multiput(535,809)(0.00000,-8.00000){18}{\line( 0,-1){  4.000}}
\put(565,709){\line( 1,-1){ 40}}
\put( 20,790){\makebox(0,0)[lb]{\raisebox{0pt}[0pt][0pt]{ \el}}}
\put(320,790){\makebox(0,0)[lb]{\raisebox{0pt}[0pt][0pt]{ \blrbar}}}
\put(150,755){\makebox(0,0)[lb]{\raisebox{0pt}[0pt][0pt]{   Z}}}
\put(135,625){\makebox(0,0)[lb]{\raisebox{0pt}[0pt][0pt]{       Fig 1a}}}
\put(480,734){\makebox(0,0)[lb]{\raisebox{0pt}[0pt][0pt]{ \rm x}}}
\put(580,734){\makebox(0,0)[lb]{\raisebox{0pt}[0pt][0pt]{\rm y}}}
\put(495,625){\makebox(0,0)[lb]{\raisebox{0pt}[0pt][0pt]{ Fig 1b}}}
\put(440,790){\makebox(0,0)[lb]{\raisebox{0pt}[0pt][0pt]{ \el}}}
\put(615,790){\makebox(0,0)[lb]{\raisebox{0pt}[0pt][0pt]{ \blrbar}}}
\put( 20,650){\makebox(0,0)[lb]{\raisebox{0pt}[0pt][0pt]{ \elbar}}}
\put(320,650){\makebox(0,0)[lb]{\raisebox{0pt}[0pt][0pt]{ \blr}}}
\put(450,650){\makebox(0,0)[lb]{\raisebox{0pt}[0pt][0pt]{ \elbar}}}
\put(615,650){\makebox(0,0)[lb]{\raisebox{0pt}[0pt][0pt]{ \blr}}}
\end{picture}

\end{center}
\vskip 36pt

Let us consider the first case. It should be noticed that the matrix elements
$\delta
\bar{M}_{L,R} $  (we call $\delta \bar{M}_{L,R}$   the matrix element
$\delta M_{L,R}$ after imposing  the condition
eq(\ref{eq:relat1}))   cannot be generated  by new physics contributing a
Feynman
diagram with the $Z$  coupled to fermions through a vertex loop  diagram (Fig
1a). The measurement of the polarisation asymmetry at SLAC is sensitive only to
modifications of the electron vertex as in Fig. 1a. As discussed above such a
non-standard  $Z_0$-electron  effective vertex coupling  introduced to
explain the $A_{LR}^{SLAC}$ would necessarily increase the $A_{FB}^e$ measured
at LEP. (This conclusion applies even if one adds a box diagram contribution as
well as its correction is aalso lways positive, as
it is evident from eq(\ref{eq:interf}) (with the index $q\rightarrow e$.)) Thus
we conclude
mechanism i). cannot reconcile LEP and SLAC.

The second possibility is that the interference is due to the imaginary part of
a box
diagram involving states beyond the Standard Model as in Fig 1b.

 However a more detailed analysis of the conditions (\ref{eq:relat1}) shows
that such a
term  does {\it not} introduce a matrix element of the type $\delta
\bar{M}_{L,R}$. We
know the value
$g_V^e/g_A^e=1-4 \Sw$ and $g_V^b/g_A^b=1-4/3 \Sw$ within the Standard Model
(for
simplicity we have taken $q=$bottom; the conclusions are the same for the other
quarks). Using this in eq(\ref{eq:relat1}) implies
$|g_A^{b\prime}|>|g_V^{b\prime}|$.
This means that the imaginary part of the box diagrams  $\delta \bar{M}_{L,R}$
of the
two processes  \ba
\label{eq:process1}
e_L+\bar e_L &\longrightarrow& b_L +\bar b_L \\
e_L+\bar e_L &\longrightarrow&  b_R +\bar b_R \nonumber\\  \nonumber\ea
must have opposite signs.
To see that this is not possible note that the only terms that can change sign
between the
two processes come from the
propagators X and Y (the remainder of the graph comes in two complex conjugate
parts
with a definite sign). However these propagators are both spacelike and hence
the sign
is independent of the identity of the states X and Y which may change for the
two
processes of eq(\ref{eq:process1}). Thus we see the sign of the box diagrams
of the two
processes (\ref{eq:process1}) must be the same and they can never give  the
correct
$|g_A^{b\prime}|>|g_V^{b\prime}|$. (The argument may be
generalised to more complicated higher loop graphs.)

Thus we are left with option iii). as the only possible source of a matrix
element of the
type  $\delta M_{1}$ is the exchange of a new gauge boson, $Z'$. Provided it is
produced
nearly on resonance its amplitude will be largely imaginary as desired. Unlike
the box
graphs in this case can one have opposite signs for $\delta\sigma_L$ and
$\delta
\sigma_R$ simply through the choice of the $Z'$ couplings. In the next section
we
consider in detail whether such a new
contribution can indeed explain the discrepancy between LEP and SLAC.

\section*{Numerical analysis}
Here we consider the couplings of the $Z'$ needed to change the peak
observables.
As stressed above its coupling to the electron must be small compared to the
$Z$   in
the channel $e^+ + e^-\rightarrow e^++e^-$. On the other hand we need
an measureable contribution  in the channel  $e^+ + e^-\rightarrow q+\bar q$,
so we
need a sizeable coupling  to a quark. We start by assuming the new $Z'$ couples
only to
the b (and t quarks). Including (small) non-interference effects there are
three
experimental measurements sensitive to such a new $Z'$
contribution namely  $A_{LR}^{SLAC}=0.163\pm0.0079$,
$R_b=0.2192\pm0.0018$ and $A_{FB}^b=0.0967\pm0.0038$
\cite{glasgow}\cite{altlampe}.
These are determined by the three independent parameters\footnote{The
other parameters can always be reabsorbed into a redefinition of this three
ones.}
$g_A^{e\prime}/g_V^{e\prime}$,$g_V^{b\prime}/g_A^{b\prime}$ and  $\delta
a$\footnote{$\delta a$ and $a$  are  defined setting $g_V^{e\prime}=1/2$ and
$g_A^{b\prime}=1/2$, and $g_A^{e}=g_A^{b}=1/2$ }.   A fit (with
$M_{top}=175~GeV$ and $M_{Higgs}=300~GeV$) gives\footnote{Here we
assume that all the $R_b$ data come only from $Z_0$-peak. Including off peak
data of
$R_b$  needs the knowledge  of both the $Z_0$ and  $Z^\prime$ lineshapes.  }
\ba
\label{eq:fit}
{g_A^{e\prime}\over g_V^{e\prime}}=-0.046\pm0.18 \nonumber\\
{g_V^{b\prime}\over
g_A^{b\prime}}=-0.716\pm0.03  \nonumber \\ {\delta a\over a} =-0.13\pm0.05.
\ea
Here we have arranged that the new neutral $Z'$ simultaneously explains  the
discrepancy
between LEP and SLAC and the small excess over the Standard Model prediction in
$R_b$. Since the values of eq(\ref{eq:fit}) nearly satisfy eq(\ref{eq:relat1})
the contribution of the interference term in eq(\ref{eq:int}) is comparable to
the
non-interference term and the discrepancy between the predicted and measured
values
of $R_b$ is of the minimum order to be expected from the need to explain the
discrepancy between the SLAC and CERN results. However this expectation is not
absolute as it is possible to find a fit consistent with the
Standard Model result for $R_b$ by fixing $g_A^{e\prime}/g_V^{e\prime}$ and
$g_V^{b\prime}/g_A^{b\prime}$ so the discrepancy is not a firm prediction of
the new
neutral current.

\section*{Tests of a new neutral gauge boson}
We have shown that the SLAC and LEP results may only be
reconciled through a new $Z'$ gauge boson nearly degenerate with the $Z$. In
practice this means that the $Z'$ should lie within $\Gamma_Z+\Gamma_{Z'}$ of
the Z mass. The most obvious test of this possibility will be forthcoming when
SLAC measure the asymmetry off the $Z$ peak for only in the case of exact
degeneracy of the $Z$ and $Z'$ will the line shape remain unchanged. What about
further tests?  By construction the most significant effects have been
put in the b quark sector and we have seen that this can lead to observable
deviations
from the Standard Model. However this depends on the precise choice of
$g_A^{e\prime}/g_V^{e\prime}$ and
$g_V^{b\prime}/g_A^{b\prime}$ and does not provide a definitive test.

However Eq(\ref{eq:fit}) strongly constrains the relative $Z'$ couplings to the
electron channel. The matrix elements
$|M_0|^2$ and $|\delta M_1|^2$ are given by \be
|M_0|^2\sim{\Gamma_e \Gamma_b\over \Gamma_Z^2}~~~;~~~ |\delta
M_1|^2\sim{\Gamma_e^\prime \Gamma_b^\prime\over
\Gamma_{Z^\prime}^2} \ee
which allow us to estimate\footnote{Here we assume the dominance of the
$b-$channel
$\Gamma_b^\prime/\Gamma_{Z^\prime}\simeq1$.} \be
\label{eq:ratio}
{\Gamma_e^\prime\over \Gamma_b^\prime}\simeq 0.8\times 10^{-4}. \ee

As we have already noted even if the coupling  of the $Z^\prime$ to the
electron is very
small, because the interference between the matrix elements of $Z$ and
$Z^\prime $
does not vanish when both final and initial states are electrons, there may be
significant
effects in this channel.  We can use the fit
(\ref{eq:fit}) and the ratio (\ref{eq:ratio}) to predict the effects to the
$e^++e^-
\rightarrow e^++e^-$ channel. While the interference does not affect the total
cross
section (this may be seen to follow from the predominantly vector-like nature
of the new
current to the electron) the effect to the forward-backward asymmetry is  \be
\label{eq:prediction}
\delta A_{FB}^e=0.004^{+0.0039}_{-0.0026}.
\ee
Thus a $Z'$ coupled principally to the b quark can be detected from a precision
measurement of the forward-backward asymmetry of the electron (in the case of
lepton universality also the forward-backward asymmetries of the $\tau$ and the
$\mu$  are similarly affected\footnote{We note that the current experimental
value
(with $M_{top}=174~GeV$ and $M_{Higgs}=300$)
is $\delta A_{FB}^e=0.0\pm0.0034$  and  $\delta A_{FB}^l=0.002\pm0.0016$ (with
lepton
universality).}, while there are negligible  effects in the $\tau$-polarization
 measurements).

Of course the fit of eq(\ref{eq:fit}) depends on the assumption that only the
$b,\bar b$
final states are affected by this $Z^\prime$. If we assume that more quarks are
(equally)
coupled to this new gauge boson the electron coupling will be reduced by the
number
of quark couplings assumed\footnote{Note that the absolute magnitude of the
$Z'$ coupling is not determined as we have assumed it to be produced on
resonance.}. In this case the effects on $R_b$, $A_{FB}^b$ and
$A_{FB}^e$ will similarly be reduced. It will also lead to an excess of two jet
production near the $Z_0$ mass in the  $p$-$\bar p$  colliders (UA2 gives
$\sigma=9.6\pm2.3(stat)\pm1.1(syst)~nb$ which is only slightly above the
Standard model prediction ($5.8~nb$) \cite{ua2} - this constrains the $Z'$
couplings to light quarks in this case to be comparable to the $Z$ couplings.).
Similarly a $Z'$ coupling to top quarks will also enhance $t \bar{t}$
production, again going in the direction favoured by current experimental
measurements.

We conclude that the left-right asymmetry measurement of SLAC is compatible
with all
LEP measurements {\it only} if we assume the existence of a $Z^\prime$ with
resonant contribution which overlaps the
$Z_0$ lineshape\footnote{We have deliberately avoided adding any theoretical
prejudice on the possible nature of the new physics but it must be admitted
that it is difficult to provide a convincing theoretical argument leading to
such a degeneracy in $Z$, $Z'$ masses.}. Signals of such a gauge boson could
come from small deviations from the Standard Model predictions for $R_b$,
$A_{FB}^e$ or from larger deviations in hadron colliders giving enhanced two
jet production (if the $Z'$ is coupled to light quarks) or enhanced $t \bar{t}$
production (if the $Z'$ is coupled to top quarks). Further tests are available
at SLAC for the prediction is that $R_b$ should change (if the $Z'$ is coupled
to b quarks) for different initial electron polarization and, more
definitively, the polarised line shape should vary due to the different
interference pattern expected off resonance.

\section*{Acknowledgements}
We  would like to thank G. Altarelli, S.Leone and M.Mangano for helpful
discussions.

\end{document}